\documentstyle[sprocl]{article}

\input{psfig} 
\def\be{\begin{equation}}
\def\ee{\end{equation}}
\def\bea{\begin{eqnarray}}
\def\eea{\end{eqnarray}}

\def\shiftleft#1{#1\llap{#1\hskip 0.04em}}
\def\shiftdown#1{#1\llap{\lower.04ex\hbox{#1}}}
\def\thick#1{\shiftdown{\shiftleft{#1}}}    
\def\b#1{\thick{\hbox{$#1$}}}

\begin{document}

\title{EXCHANGE CURRENTS IN RADIATIVE HYPERON DECAYS}
\author{GEORG WAGNER}
\address{Centre for the Subatomic Structure of Matter (CSSM)
         University of Adelaide, Australia 5005} 

\maketitle
\abstracts{A short overview of motivations and successes of
           two-body exchange currents between constituent 
           quarks for electromagnetic hadron observables like 
           charge radii, magnetic and quadrupole moments is given.
           We then predict and analyze exchange current effects on the 
           radiative 
           decay widths of decuplet hyperons, which are to be measured soon. 
           In our chiral constituent quark model, exchange currents 
           dominate the $E2$ transition amplitude, 
           while they largely cancel for the $M1$ transition amplitude. 
           Strangeness suppression of the radiative hyperon decays is 
           weakened by exchange currents.
           The SU$_3$(F) flavor symmetry breaking for the negatively
           charged hyperons is strong.}

\section{Introduction}
\label{sec:intro}
The naive quark model successfully describes baryon magnetic moments.
Assuming isospin symmetry $\mu_u=-2\mu_d$, without parameters, 
the proton to neutron magnetic moment ratio -3/2 is close to 
the experimental value:
\begin{equation}
  \mu_p=\frac{4\mu_u-\mu_d}{3}  \; ,\; \mu_n=\frac{4\mu_d-\mu_u}{3}  
  \Rightarrow \frac{\mu_p}{\mu_n} = -\frac{3}{2} \simeq\frac{2.793}{-1.913}
\label{eq:impmom}
\end{equation}
In this impulse approximation -- sketched in Figure
\ref{figure:currents}(a) -- the photon probing the hadron
is absorbed on a single quark, with the two other quarks acting as spectators. 
If this picture were correct, observables like the neutron charge 
radius or the quadrupole moment of the nucleon or the $\Delta$ could only be 
described by d-state deformations in the hadron wave functions.
Deformations arise from residual tensor interactions between quarks. 
These arguments led to recently performed measurements \cite{Ppp95} of the
photo-production of the $\Delta$-resonance off the nucleon, by
which one wishes to extract informations on the deformations of the nucleon 
and/or $\Delta$ by the electric quadrupole to magnetic dipole ratio of the 
transition amplitudes.

In contrast to the above ideas, the baryon mass spectrum 
gives no evidence of strong tensor forces between quarks.
 Furthermore, in low-energy QCD the quasi-particle constituent 
quarks are strongly correlated, interacting for 
example via the pseudoscalar meson cloud surrounding them.
The impulse approximation violates the continuity equation for the
electromagnetic current and two-body exchange currents, where the
photon momentum is distributed among constituent quarks via 
their interactions 
(sketched in Figures \ref{figure:currents}(b) -- \ref{figure:currents}(e)), 
have to be consistently introduced.

In recent years, several works \cite{Buc91,Wag95,Buc97,Mey97} have studied 
the effect of two-body exchange currents on electromagnetic baryon observables.
A first important observable for exchange currents is the negative square 
charge radius of the neutron \cite{Kop95} $r_n^2 = -.113(3)$ fm$^2$.
Within different models \cite{Buc91,Chr96,Dlu98}, its value 
has been clearly assigned to the non-valence degrees of freedom in the nucleon.
One of these models, namely the concept of exchange currents \cite{Buc91}, 
allows to derive an analytic relation for $r_n^2$ and the $\Delta$-nucleon 
mass splitting which is due to the residual spin-dependent interactions 
between quarks
\begin{equation}
  r_n^2 = -\frac{M_\Delta -M_N}{M_N} \, b_N^2 \; .
\label{eq:alf}
\end{equation} 
It is fulfilled for a quark core size $b_N$=0.61 fm, which is the only model
parameter in Eq.\  (\ref{eq:alf}).
The intimate connection of electromagnetic and hadronic observables in Eq.\ 
(\ref{eq:alf}) is due to the continuity equation for the electromagnetic 
current (see below, Eq.\ (\ref{eq:conti})) which relates the two-body 
interactions in the quark model Hamiltonian
and the two-body components (exchange currents) in the 
current operator \footnote{In impulse
approximation in a configuration mixing calculation with d-state admixtures, 
one needs a much to big quark core size of $b_N\sim$ 1 fm to describe $r_n^2$. 
Using a value of $b_N\sim$ 0.6 fm as required for other hadronic 
observables, one obtains a much to small neutron charge radius of 
$r_n^2 = -.03$ fm$^2$.}.

In this contribution we discuss where exchange currents are important and 
why the magnetic moment results sketched in Eq.\ (\ref{eq:impmom}) remain valid
after  introduction of exchange currents \cite{Wag95}. 
We then analyze exchange current effects in 
the radiative hyperon decays \cite{Wag98}.
We demonstrate the possible dominance of exchange currents in 
the $E2/M1$ ratios and study their SU$_3$(F) symmetry breaking properties.

\section{Exchange currents in the chiral constituent quark model}
\label{sec:model}
\subsection{Hamiltonian, wave functions, parameters and baryon masses}
\label{subsec:ham}
Constituent quarks emerge as the effective quasi-particle degrees of freedom 
in hadron physics due to the spontaneously broken chiral symmetry of 
low-energy QCD.
A non-relativistic realisation of the chiral quark model Hamiltonian in the 
case of three quark flavors $u,d,s$ is \cite{Buc97,Wag98}
\begin{equation}
  H = \sum_{i=1}^{3} \big( m_i+ {{\bf p}^2_i\over 2m_i} \big)
  - {{\bf P}^2\over 2M} 
  + \sum_{i<j}^{3} \left( V^{\rm{Conf}}(i,j) 
                        + V^{\rm{PS}}(i,j)  
                        + V^{\rm{OGE}}(i,j)  \right)
\label{eq:ham}
\end{equation}
Here, a quadratic confinement potential $V^{\rm{Conf}}$ is used.
The radial form of the confinement potential is according to our experience 
not crucial for the discussion of hadronic ground state properties.
The pseudoscalar meson octet (PS), that constitute the Goldstone bosons of the
symmetry breaking, provide the intermediate range
 interactions between quarks $V^{\rm{PS}}$.
We use experimental PS-meson masses and one universal cut-off parameter
$\Lambda$=4.2 fm$^{-1}$ for regularisation. 
The quark-meson coupling constant is related to the pion-nucleon coupling.
The $\sigma$-meson as the chiral partner of the pion is included \cite{Wag95}
whereas the heavier scalar partners of the Kaon and $\eta$ 
are neglected.
At short range, the residual interactions comprise one-gluon exchange 
$V^{\rm{OGE}}$ in Fermi-Breit form without retardation corrections \cite{deR75}.

\vspace{-0.45cm}
\begin{table}[htb]
\caption[Hamiltonian]{Individual contributions of Hamiltonian 
        (\ref{eq:ham}) to octet and decuplet hyperon masses  
        (quark masses, kinetic energy, confinement potential, one-gluon-, 
        pseudoscalar meson- (PS) and $\sigma$-exchange potentials). 
        Experimental values \cite{Bar96} average over particles with 
        different charge. All quantities are given in [MeV]. 
\label{table:masses}}
  \vspace{0.1cm}
\begin{center}
  \footnotesize
\begin{tabular}{| l | r  r  r  r  r  r  r | r |}  \hline 
\rule[-1mm]{0mm}{4.5mm}[MeV] & $\sum_i m_i$ & Kin. & Conf. & Gluon & PS & 
$\sigma$ & Total & Exp.\cite{Bar96}                         \\[0.05cm] \hline
N             &   939 & 497 & 204 & -531 & -115 & -54 &  939 &  939 \\ 
$\Sigma$      &  1195 & 497 & 173 & -562 &  -51 & -65 & 1188 & 1193 \\ 
$\Lambda$     &  1195 & 497 & 173 & -588 &  -88 & -65 & 1124 & 1116 \\ 
$\Xi$         &  1451 & 497 & 143 & -652 &  -45 & -78 & 1316 & 1318 \\ \hline 
$\Delta$      &   939 & 497 & 204 & -326 &  -27 & -54 & 1232 & 1232 \\
$\Sigma^\ast$ &  1195 & 497 & 173 & -423 &  -18 & -65 & 1359 & 1385 \\
$\Xi^\ast$    &  1451 & 497 & 143 & -561 &  -13 & -78 & 1439 & 1535 \\
$\Omega^-$    &  1707 & 497 & 112 & -595 &  -12 & -95 & 1615 & 1672 \\ \hline 
\end{tabular}
\end{center}
\end{table}
 
\vspace{-0.25cm}
We use spherical $(0s)^3$ oscillator states \cite{Wag95} 
(containing symmetry breaking effects by the quark masses $m_s\neq m_d$) 
and SU$_{SF}$(6) spin-flavor states for the baryon wave functions.
 For chosen quark masses $m_u$=$m_N/3$=313 MeV and $m_u/m_s$=0.55,
the effective quark-gluon coupling, the confinement strength and the wave 
function oscillator parameter $b_N$ are determined from 
the baryon masses \cite{Wag98}.  
Results for the octet and decuplet ground state masses
are shown in Table \ref{table:masses}.

\vspace{-0.25cm}
\subsection{Continuity equation, electromagnetic currents, and magnetic moments}
\label{subsec:elmag}
 For illustration, use the Fourier-transformed Heisenberg equation of 
motion for the charge density 
$i\partial/\partial t\, \rho (\vec{x},t) = \left[ H , \rho (\vec{x},t) \right]$
in the continuity equation for the electromagnetic current 
$\partial_\mu j^\mu =0$:
\begin{equation}
  \vec{q} \cdot \vec{j}(\vec{q}) = \left[ H , \rho (\vec{q}) \right] \quad .
\label{eq:conti}
\end{equation}
Eq.\ (\ref{eq:conti}) shows, that for point charges 
$\rho (\vec{q}) = e_i \, \exp (i\vec{q}\cdot\vec{r}_i)$ interacting 
by a Hamiltonian $H$, momentum- and/or flavor-dependent interactions
in $H$ have their corresponding electromagnetic 
two-body current operators $\vec{j}$.
 
\begin{figure}[htb]
  \psfig{figure=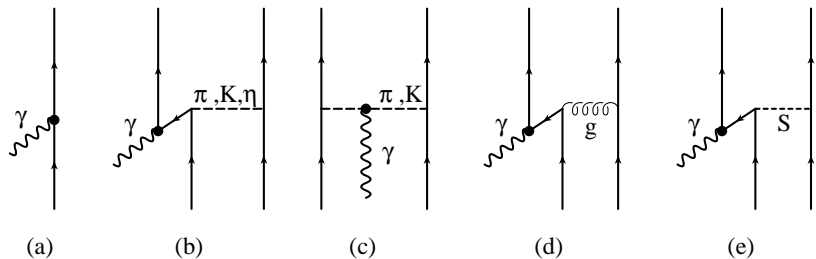,height=1.35in}
\caption{Electromagnetic one- and two-body currents as required by the 
         continuity equation.
         (a) Impulse approximation, (b) PS-meson pair current 
	 ($\pi , K, \eta$), (c) PS-meson in-flight current ($\pi , K$), 
	 (d) gluon-pair current, and (e) scalar exchange current 
	 (confinement and $\sigma$-exchange).
\label{figure:currents}}
\end{figure}

Only the longitudinal part of the three-vector current is determined by
Eq.\ (\ref{eq:conti}).
The electromagnetic currents to be included for Hamiltonian (\ref{eq:ham}) 
are either constructed by minimal substitution or like in the present
case by non-relativistic reduction of the relevant Feynman diagrams 
\cite{Buc97} shown in Figure \ref{figure:currents}.
The extension of the exchange current operators
to three quark flavors can be found in \cite{Wag95,Wag98}.
The spatial components for example
satisfy the continuity equation (\ref{eq:conti}) 
with Hamiltonian (\ref{eq:ham}) to order ${\cal{O}}(1/m_q)$.

\vspace{-0.1cm}

\vspace{-0.2cm}
Results for the magnetic moments of the octet baryons \cite{Wag95}
$\mu_B \propto \langle B\vert j^+ \vert B \rangle$
are given in Table \ref{table:octet}.
Individual exchange current contributions can be as large as
40$\%$ of the impulse approximation. 
We observe substantial cancellations between the gluon-pair- and 
the scalar-pair-currents (confinement and one-sigma-exchange).
Second, partial cancellations between the PS-meson in-flight and 
the PS-meson pair term occur.
The PS-mesons play an important role, because they reduce the strong 
quark-gluon coupling (adding to the $\Delta$-nucleon mass splitting)
and thus the gluon exchange current contribution.

\vspace{-0.4cm}
\begin{table}[htb]
\caption[octet magnetic moments]{Octet baryon magnetic moments. 
	The impulse and the exchange current contributions
	are listed separately: PS-meson-pair  and PS-meson-in-flight,
	gluon-pair, confinement- and $\sigma$-pair currents.
	All quantities are given in nuclear magnetons 
	${\mu}_N$= ${e\over 2M_N}$.
	Experimental numbers are from \cite{Bar96}.
\label{table:octet}}
  \vspace{0.1cm}
\begin{center}
  \footnotesize
\begin{tabular}{| l | r  r  r  r  r  r  r | r |}\hline
\rule[-1mm]{0mm}{4.5mm}$[\mu_N]$ &
Imp. & PS-pair &-in-flight & Gluon & Conf. & $\sigma$ & Tot. & Exp.\cite{Bar96}
\\[0.05cm] \hline
   p       &  3.00 & -0.13 & 0.41 & 0.62 &-1.30 & 0.35 & 2.94& 2.793	     \\
   n       & -2.00 &  0.19 &-0.41 &-0.21 & 0.87 &-0.23 &-1.79&-1.913	     \\
$\Sigma^+$ &  2.85 & -0.00 & 0.07 & 0.80 &-1.06 & 0.37 & 3.02& 2.458$\pm$.01 \\
$\Sigma^0$ &  0.85 & -0.02 & 0.03 & 0.17 &-0.29 & 0.10 & 0.84&		     \\
$\Sigma^-$ & -1.15 & -0.05 & 0    &-0.47 & 0.48 &-0.16 &-1.35&-1.160$\pm$.025\\
$\Lambda$  & -0.55 &  0.03 &-0.03 & 0.03 & 0.10 &-0.05 &-0.46&-0.613$\pm$.004\\
$\Lambda\!\leftrightarrow\!\Sigma^0$ 
           &  1.73 & -0.15 & 0.26 & 0.26 &-0.67 & 0.23 & 1.67& 1.610$\pm$.08 \\
$\Xi^0$    & -1.40 &  0.08 &-0.07 &-0.10 & 0.34 &-0.17 &-1.32&-1.250$\pm$.014\\
$\Xi^-$    & -0.40 & -0.02 & 0    &-0.13 &0.004 &-0.02 &-0.55&-0.651$\pm$.003 
\\ \hline
\end{tabular}
\end{center}
\end{table}

Overall, exchange currents provide less than 10$\%$ corrections to the 
magnetic moments.
The cancellations are most pronounced in the S=--1 sector, and we observe 
a nice improvement for the $\Xi$'s, in particular we obtain 
$\vert\mu_{\Xi^-}\vert > \vert\mu_\Lambda\vert$ in agreement with experiment.
Magnetic moments being very sensitive to the quark core size $b_N$, 
one obtains the strongest cancellations and best agreement 
with experimental data
for a core size $b_N\sim$ 0.6 fm in accordance with Eq.\ (\ref{eq:alf}).
 
\subsection{Interrelations between observables}
\label{subsec:inter}
Within the present approach, the $\Delta$ charge radius is larger than the 
proton radius by the amount of the neutron radius \cite{Buc97}, i.e.\ it is
given by the isovector charge radius of the nucleon: $r_\Delta^2=r_p^2-r_n^2$.
Even more interesting is the analytic result for the quadrupole moment of the
$\Delta^+$ and the $C2$ ($E2$) multipole amplitude in the
$\gamma N\leftrightarrow \Delta$ transition \cite{Buc97}: 
\begin{equation}
  Q_{\Delta^+} = \sqrt{2}\cdot Q_{\gamma N\leftrightarrow\Delta}       
               = r_n^2 
               = -b_N^2\;\frac{m_\Delta-m_N}{m_N}  \qquad . 
\label{eq:rel}
\end{equation}
Quark model calculations using D-state admixtures obtain values three to four
times smaller, while the exchange current result 
of Eq.\ (\ref{eq:rel}) gives the correct empirical \cite{Bar96} quadrupole 
transition amplitude $\left( r_n^2\right)^{\rm{exp}} = 
 1.5 \cdot Q_{\gamma N\leftrightarrow\Delta}^{\rm{exp}}$.
Again, using $(0s)^3$ wave functions, we connect and explain observables 
by the same non-valence degrees of freedom, the spin-dependent interactions
between constituent quarks.
At the prize of using simplifying model dynamics, we obtain a qualitative
understanding of experimental facts, even including a good agreement with data.

\begin{figure}[htb]
  \psfig{figure=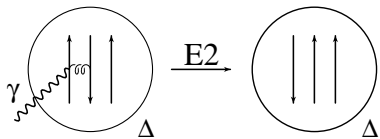,height=0.75in} 
\caption{Two-body spin flip induced by gluonic and PS-meson exchange currents.
\label{figure:flip}}
\end{figure}

In Eq.\ (\ref{eq:rel}), we used Siegert's theorem in the long wavelength limit,
which connects $C2$ and $E2$ amplitudes and allows to calculate the $E2$ 
form factor 
at small momentum transfers from the charge density $\rho ({\bf{q}})$.
Technically, only the gluon- and PS-meson-pair charge density operators 
can contribute for spherical wave functions.
Their tensorial structure in spin-space allows for a double spin-flip of the
two participating quarks ${\b{\sigma}_i^+}{\b{\sigma}_j^-}$ as the only
mechanism by which an $E2$ (or $C2$) photon can be absorbed \cite{Buc97}.
This process is sketched in Figure \ref{figure:flip} for the case of the
$\Delta$ quadrupole moment (for a $\Delta$ with S,S$_z$=3/2,1/2).

\section{Radiative hyperon decays}
\label{sec:decay}
Current experimental programs aim at a detailed measurement of the radiative
decays of some $\Sigma^{\ast}$ and $\Xi^{\ast}$ hyperons \cite{Rus95}.
Many model calculations have been performed.
Besides the pioneering study of Lipkin and quark model impulse approximation 
predictions \cite{Lip73} (there, neglecting exchange currents and 
$D$-state admixtures, all decays are pure $M1$ transitions),
hyperon decays have been studied for example in  SU$_F$(3) Skyrme 
models \cite{Aba96}, chiral bag models \cite{Dlu98}, heavy baryon chiral 
perturbation theory\cite{But93}, 
or in a quenched lattice calculation\cite{Lei93}.

The reasons for increased interest in these observables are twofold.
Certainly, as for $\gamma N\leftrightarrow\Delta$, the $E2/M1$ ratios 
in the radiative hyperon decays contain information on deformations, if not of 
the valence quarks, then of the non-valence-quark distributions in the baryons.
Comparison of model predictions with experiment may provide another signal of 
exchange currents and may pin down the importance of vector (gluon) vs.\ 
pseudoscalar degrees of freedom in the effective quark-quark interaction.

Second, the decays are sensitive to SU$_F$(3) flavor symmetry breaking.
The decay widths of the negatively charged hyperons
$\Sigma^{\ast -}\rightarrow\gamma\Sigma^-$ and
$\Xi^{\ast-}\rightarrow\gamma\Xi^-$
are zero in a SU$_F$(3) flavor-symmetric world.
It has been speculated \cite{Aba96} that these decays remain almost
forbidden even after SU$_F$(3) symmetry breaking.
Strangeness suppression, i.\ e.\ the decrease of the decay
amplitude with increasing strangeness of the hyperon, is best studied 
comparing transitions involving wave functions which are identical except for 
the replacement of d- by s-quarks, like $\gamma N\leftrightarrow \Delta^{0}$ 
and $\gamma \Xi^{0}\leftrightarrow \Xi^{\ast 0}$.

\subsection{Magnetic dipole transition amplitudes}
\label{subsec:mag}
The various contributions to the $M1$ and $E2$ (using Siegert's theorem)
transition moments defined in \cite{Buc97,Wag98} 
are given in Tables \ref{table:mamo} and \ref{table:quamo}.
Individual exchange current contributions to the $M1$ moments can be 
as large as 60$\%$ of the impulse approximation. 
Like for the octet baryon magnetic moments \cite{Wag95} in 
Table \ref{table:octet}, substantial cancellations between  
gluon-pair- and   scalar-pair-currents (confinement and one-sigma-exchange) 
occur for all decays.
Due to partial cancellations between the PS-meson in-flight and the PS-meson 
pair term, the total PS-meson contribution to the $M1$-amplitude is again small.

Exchange currents thus provide less than 10$\%$ overall corrections to the 
transition magnetic moments.
However, some striking systematics are observed.
Strangeness suppression in impulse approximation, i.\ e.\ in a picture of
valence quarks only, is considerable due to $m_u/m_s=$0.55--0.6
(first column in Table \ref{table:mamo}).
Exchange currents decrease the $\gamma N\leftrightarrow\Delta$ $M1$ moment,
slightly decrease the decay amplitudes in the S=--1 sector and slightly 
increase the result for the S=--2 sector.
Therefore, strangeness suppression is for all six strange decays
considerably reduced when exchange currents are 
included \footnote{In particular, the reduction of the
$\gamma\Xi^{0}\leftrightarrow\Xi^{\ast 0}$ impulse result
$\mu_{\rm{imp}}^{\gamma\Xi^{0}\leftrightarrow\Xi^{\ast 0}}$=2.404$\mu_N$ as
compared with the $\gamma n\leftrightarrow\Delta^0$ transition magnetic moment
$\mu_{\rm{imp}}^{\gamma n\leftrightarrow\Delta^0}$=2.828$\mu_N$
practically disappears when exchange currents are included, and we obtain
$\mu_{\rm{tot}}^{\gamma n\leftrightarrow\Delta^0} \simeq
 \mu_{\rm{tot}}^{\gamma\Xi^0\leftrightarrow\Xi^{\ast 0}}$=2.428$\mu_N$.}.
Strangeness suppression is mostly strong in Skyrme model calculations
\cite{Aba96}, while lattice results from \cite{Lei93} agree reasonably
well with our predictions.

\vspace{-0.25cm}
\begin{table}[htb]
\caption[M1 Transition moments]{Transition magnetic moments 
        $\mu$ of decuplet baryons.
        The impulse and the various exchange current contributions
        are listed separately, like in table \ref{table:octet}.
        All quantities are given in nuclear magnetons 
        ${\mu}_N$= ${e\over 2M_N}$.
        Experimentally known is the non-strange 
        $\Delta^+\rightarrow \gamma p$  transition magnetic moment 
        $\mu^{exp}_{\Delta^+\rightarrow p}$ = 3.58(9) $\mu_N$ \cite{Bar96}. 
\label{table:mamo}}
  \vspace{0.1cm}
\begin{center}
  \footnotesize
\begin{tabular}{| l | r  r  r  r  r  r  r |} \hline 
\rule[-1mm]{0mm}{4.5mm}$[\mu_N]$ &
Imp. & PS-pair & -in-flight & Gluon & Conf. & $\sigma$ & Tot. \\[0.05cm] \hline
$\gamma N\leftrightarrow \Delta$ &
   2.828 & -0.274 &  0.586 & 0.292 & -1.228 & 0.327 & 2.533          \\
$\gamma \Sigma^+ \leftrightarrow \Sigma^{\ast +}$   &   
   2.404 & -0.068 &  0.097 & 0.366 & -0.822 & 0.291 & 2.267          \\
$\gamma \Sigma^0 \leftrightarrow \Sigma^{\ast 0}$   & 
  -0.990 &  0.036 & -0.049 &-0.095 &  0.278 &-0.105 &-0.924          \\
$\gamma \Sigma^- \leftrightarrow \Sigma^{\ast -}$   & 
  -0.424 & -0.004 &   0    &-0.176 &  0.267 &-0.082 &-0.419          \\
$\gamma \Lambda \leftrightarrow \Sigma^{\ast 0}$    & 
   2.449 & -0.212 &  0.366 & 0.371 & -0.944 & 0.323 & 2.354          \\
$\gamma \Xi^0 \leftrightarrow \Xi^{\ast 0}$         & 
   2.404 & -0.117 &  0.097 & 0.431 & -0.716 & 0.329 & 2.428          \\
$\gamma \Xi^- \leftrightarrow \Xi^{\ast -}$         & 
  -0.424 &  0.009 &   0    &-0.190 &  0.235 &-0.090 &-0.460          \\ \hline 
\end{tabular} 
\end{center}
\end{table}

\vspace{-0.25cm}
The transition magnetic moments for the negatively charged hyperons
($\sim$--0.4$\mu_N$) deviate considerably from the SU$_F$(3) flavor-symmetric
value 0, when the quark mass ratio $m_u/m_s$=0.55 is used.

An interesting comparison can be made for the $M1$
moments of $\gamma\Sigma^{+}\leftrightarrow\Sigma^{\ast +}$ and
$\gamma\Xi^{0}\leftrightarrow\Xi^{\ast 0}$ or
$\gamma\Sigma^{-}\leftrightarrow\Sigma^{\ast -}$ and
$\gamma\Xi^{-}\leftrightarrow\Xi^{\ast -}$.
They are pairwise equal in impulse approximation
(cf.\ Table \ref{table:mamo}), and would 
also be equal after inclusion of exchange currents
if SU$_F$(3) flavor symmetry was exact.
Gluon- and scalar-exchange currents lead to 
deviations of the order of 10$\%$.
Other model calculations differ qualitatively (both in signs and magnitudes)
from our prediction for the SU$_F$(3) violation of these sum rules.

\subsection{$E2$ transition amplitudes, $E2/M1$ ratios and decay widths}
\label{subsec:ele}
It seems that the $M1$ transition moments should be measured to very high 
accuracy if one wishes to discriminate between models or to establish clear 
signals for exchange currents.
 Furthermore, most approaches underestimate the only empirically known 
transition magnetic 
moment $\mu^{\rm{exp}}_{\Delta^+\rightarrow p}$=3.58(9) $\mu_N$ and decay width 
$\Gamma^{\rm{exp}}_{\Delta\rightarrow\gamma N}$=610--730 keV \cite{Bar96}.
This problem has not been solved by the exchange currents included here.
More promising are the $E2$ transition amplitudes.
Our calculated $E2$ transition amplitude for $\gamma N\leftrightarrow \Delta$
of --.089 fm$^2$ agrees with the recent experimental data
\cite{Bec97,Han97} $Q^{\rm{exp}}_{\Delta^+\rightarrow p}=-0.085(13)$ fm$^2$.

The hyperon transition quadrupole moments shown in Table \ref{table:quamo} 
receive large contributions from the PS-meson and gluon-pair diagrams of 
  Fig.\ \ref{figure:currents}b  and Fig.\ \ref{figure:currents}d.
We recall that the $E2$  moments would be zero in impulse approximation 
for spherical valence quark configurations, which we used here.
$E2$ transition moments for negatively charged hyperons $\Xi^{\ast -}$ 
and $\Sigma^{\ast -}$ deviate from the SU$_F$(3) flavor-symmetric value 0.
The gluon contributes strongly to most transition quadrupole moments, on the
average $\sim$2/3 of the total $E2$ moment.
The experiments may give important hints on the relevance of effective 
gluon degrees of freedom in hadron properties.  
Our results are mostly larger in magnitude than Skyrme model results 
\cite{Aba96}, but somewhat smaller than recent lattice results\cite{Lei93}.

\vspace{-0.2cm}
\begin{table}[htb]
\caption[C2 Transition moments]{Transition quadrupole moments $Q$ of 
        decuplet baryons. 
        The gluon-pair (${Q}_{\rm{g}}$) and individual PS-meson 
        ($\pi ,K, \eta$) exchange current contributions are listed separately. 
        All transition quadrupole moments are given in [fm$^2$]. 
        The last two columns contain the radiative decay widths 
        $\Gamma$ in [keV] and $E2/M1$ ratios in [$\%$]. 
        Note that our results are given at {\bf q}$^2$=0.  
        Experimentally known is the non-strange decay width 
        $\Gamma^{exp}_{\Delta\rightarrow\gamma N}$=610--730 keV \cite{Bar96}.
\label{table:quamo}} 
  \vspace{0.1cm}
\begin{center}
  \footnotesize
\begin{tabular}{| l | r  r  r  r | r || r  r |}\hline 
\rule[-1mm]{0mm}{4.5mm}[fm$^2$] & ${Q}_{\rm{g}}$ & ${Q}_\pi$ & ${Q}_K$ & 
${Q}_\eta$ &$Q_{\rm{Tot}}$ & $\Gamma$ [keV] & $E2/M1$ [$\%$]  \\[0.05cm] \hline
$\gamma N\leftrightarrow \Delta$ &
 -.058 & -.027 &   0   & -.004 & -.089 &  350  & -3.65      \\ 
$\gamma \Sigma^+ \leftrightarrow \Sigma^{\ast +}$ &
 -.051 & -.036 &  .005 & -.009 & -.091 &  105  & -2.9       \\
$\gamma \Sigma^0 \leftrightarrow \Sigma^{\ast 0}$ & 
  .016 &  .009 &  .002 &  .002 &  .030 &  17.4 & -2.3       \\
$\gamma \Sigma^- \leftrightarrow \Sigma^{\ast -}$ & 
  .018 &  .018 & -.010 &  .006 &  .032 &  3.61 & -5.5       \\
$\gamma \Lambda \leftrightarrow \Sigma^{\ast 0}$ & 
 -.041 &   0   & -.013 &  .006 & -.047 &  265  & -2.0       \\ 
$\gamma \Xi^0 \leftrightarrow \Xi^{\ast 0}$ &
 -.035 &   0   & -.005 &  .001 & -.039 &  172  & -1.3       \\
$\gamma \Xi^- \leftrightarrow \Xi^{\ast -}$ & 
  .012 &   0   &  .010 & -.006 &  .016 &  6.18 & -2.8       \\ \hline 
\end{tabular} 
\end{center}
\end{table}

The decay width $\Gamma \propto \vert A_{3/2} \vert^2 + \vert A_{1/2} \vert^2$
is related to the helicity amplitudes $A_{3/2}$ and $A_{1/2}$, which can be 
expressed as linear combinations of the $M1$ and $E2$ transition formfactors 
\cite{Buc97,Gia90}.
The $E2/M1$ ratio of the transition amplitudes is commonly defined as 
$E2/M1 = \frac{\omega M_N}{6} Q/\mu$, where the resonance frequency
$\omega$ is given in the c.m.\ system of the decaying hyperon.
In Table \ref{table:quamo}, both observables are shown in the last two columns.
Due to cancellations of exchange current contributions
to the $M1$ transition amplitude and the relative smallness of the $E2$ 
amplitude,
the decay widths $\Gamma$ are dominated by the $M1$ impulse approximation,
i.\ e.\ by the simple additive quark model picture with valence quarks only.
Only restricted informations on non-valence quark effects should 
be expected from the experiments here, 
similar to the situation for the octet magnetic moments.

All model calculations yield large (the largest) $E2/M1$ ratios for 
negatively charged states.
However, there are important differences.
The $E2/M1$ ratio for the decays of negatively charged hyperons are 
particularly  model dependent\cite{Aba96} due to the smallness of both 
the $E2$ and $M1$ contributions.
Similarly, the $\gamma\Sigma^{0}\leftrightarrow\Sigma^{\ast 0}$ $E2/M1$ ratio 
in the  Skyrme model approaches is zero \cite{Aba96}, or almost zero,
while the SU$_F$(3) symmetry breaking and the gluon-pair current in our model 
allow a sizeable $E2/M1$ ratio of $-2.3\%$.

\section{Summary}
\label{sec:summa}
We have given a brief overview on the concept of exchange currents and 
predicted and discussed exchange current effects on the radiative decays of 
decuplet hyperons, which are to be measured soon.
In quark potential model descriptions of hadron properties, 
two-body exchange currents are necesssary to satisfy the continuity
equation for the electromagnetic current.
Exchange currents naturally explain  the negative mean square
neutron charge radius. 
Additive magnetic moments results (impulse approximation) 
remain valid since exchange currents provide only 10$\%$ corrections due to
cancellations.
Observables which are dominated by non-valence d.o.f.\ like
$r_n^2, Q_\Delta, Q_{\gamma N\!\leftrightarrow\!\Delta}$ 
are well described by exchange currents, and simple analytic relations 
between observables can be derived.
The exchange current concept helps to gain qualitative insight in physical 
origins of hadron phenomenology.

The widths of the radiative hyperon decays are determined by the 
impulse approximation $M1$ transition, due to cancellation effects for the 
$M1$ exchange current contributions and the smallness of the $E2$ transition.
In contrast, exchange currents dominate the $E2/M1$ ratios, where the gluon- 
and PS-meson-pair charge densities lead to non-zero $E2$ amplitudes for all 
hyperon decays.
Experimental results on the $E2/M1$ ratios may provide a good test for the 
relative importance of effective gluon versus pseudoscalar degrees of freedom 
in low-energy QCD.

Individual $M1$ and in particular $E2$ transition amplitudes are sensitive to 
SU$_F$(3) flavor symmetry breaking and allow to discriminate between models.
Violations of impulse approximation or SU$_F$(3) flavor symmetry sum-rules 
are strongly model dependent. 
In our calculation, strangeness suppression is weakened by exchange currents.
The decay widths of the negatively charged hyperons deviate considerably 
from the SU$_F$(3) flavor symmetric value.

\section*{Acknowledgments}
The work reported here has been performed in collaboration with 
A.\ J.\ Buchmann and Amand Faessler (University of T\"{u}bingen, Germany).
I thank the Deutsche Forschungsgemeinschaft (DFG) for a postdoctoral 
fellowship (contract number WA1147/1-1).

\section*{References}
\def\Journal#1#2#3#4{{#1}{\bf #2}, #3 (#4)}
\def\NPA{{\em Nucl.\ Phys.\ }{\bf A}}
\def\PLB{{\em Phys.\ Lett.\ } {\bf B}}
\def\PRL{{\em Phys.\ Rev.\ Lett.\ }} 
\def\PRC{{\em Phys.\ Rev.\ }{\bf C}}
\def\PRD{{\em Phys.\ Rev.\ }{\bf D}}

\end{document}